\documentclass[twocolumn,article,superscriptaddress,showpacs]{revtex4-1}
\usepackage[utf8]{inputenc}
\usepackage{color}
\usepackage{amsmath}
\usepackage{amssymb}
\usepackage{graphicx}
\usepackage{esint}
\usepackage[unicode=true,
 bookmarks=false,
 breaklinks=true,pdfborder={0 0 0},backref=false,colorlinks=true]
 {hyperref}

\usepackage{epstopdf}\usepackage{subfigure}
\usepackage{float}
\usepackage{amsfonts}
\usepackage{bm}
\usepackage{subfigure}
\usepackage{braket}
\usepackage{braket}
\usepackage{physics}
\usepackage{xcolor}

\makeatother

\begin{document}

\title{Parallel and anti-parallel helical surface states for topological semimetals}


\author{Tiantian Zhang}
\affiliation{Department of Physics, Tokyo Institute of Technology, Ookayama, Meguro-ku, Tokyo 152-8551, Japan}
\affiliation{Tokodai Institute for Element Strategy, Tokyo Institute of Technology, Nagatsuta, Midori-ku, Yokohama, Kanagawa 226-8503, Japan}

\author{Shuichi Murakami}
\affiliation{Department of Physics, Tokyo Institute of Technology, Ookayama, Meguro-ku, Tokyo 152-8551, Japan}
\affiliation{Tokodai Institute for Element Strategy, Tokyo Institute of Technology, Nagatsuta, Midori-ku, Yokohama, Kanagawa 226-8503, Japan}
%

%
%
%
\def\LBO{Li$_{2}$B$_{4}$O$_{7}$}
\def\ZZ{$Z_{2}$}
\def\Z{$Z$}
\def\PT{$\mathcal{PT}$}
\def\GT{$\mathcal{GT}$}
\def\TG{$\mathcal{TG}$}
\def\T{$\mathcal{T}$}
\def\G{$\mathcal{G}$}
\def\P{$\mathcal{P}$}
\def\Q{$\mathcal{Q}$}
\def\C{$\mathcal{C}$}
\def\Th{$\tilde{\Theta}$}

%

\begin{abstract}

Weyl points, carrying a Z-type monopole charge \C{}, have bulk-surface correspondence (BSC) associated with 
helical surface states~(HSSs). 
When $|$\C{}$|$ $>1$, multi-HSSs can appear in a parallel manner. 
However, when a pair of Weyl points carrying \C{} $=\pm1$ meet, a Dirac point carrying \C{} = 0 can be obtained and the BSC vanishes. 
Nonetheless, a recent study in Ref.~[arXiv:2201.03238] shows that a new BSC can survive for Dirac points when the system has time-reversal (\T)-glide (\G) symmetry (\Th=\TG), i.e., anti-parallel double/quad-HSSs associated with a new \ZZ{}-type monopole charge \Q{} appear. 
In this paper, we systematically review and discuss both the parallel and anti-parallel multi-HSSs for Weyl and Dirac points, carrying two different kinds of monopole charges. Two material examples are offered to understand the whole configuration of multi-HSSs. 
One carries the Z-type monopole charge \C{}, showing both local and global topology for three kinds of Weyl points, and it leads to parallel multi-HSSs. The other carries the \ZZ{}-type monopole charge \Q{}, only showing the global topology for \Th-invariant Dirac points, and it is accompanied by antiparallel multi-HSSs. 
These diverse topological surfaces states not only offer the local and global topology for the bulk degeneracy, but also offer platforms to various topological phases.

\end{abstract}

\flushbottom
\maketitle

\thispagestyle{empty}


\section*{Introduction}

Topological semimetals are rich in categories due to the diversity of their topological degeneracies~\cite{burkov2011topological,turner2013beyond,chiu2014classification,burkov2016topological,fang2016topological,chiu2016classification,bansil2016colloquium,weng2016topological,yan2017nodal,yan2017topological,murakami2017emergence,yang2018symmetry,schoop2018chemical}. 
Topological equivalence of two semimetals can be diagnosed by either the type or the value of their topological invariants, which are quantum numbers defined uniquely for different systems, and associated with different topological surface states due to bulk-surface correspondence (BSC). 
For example, topological invariant defined for nodal-line/ring semimetals is the Berry phase associated with drumhead surface states~\cite{burkov2011topological,fang2016topological,bian2016drumhead,chan20163,deng2019nodal}, topological invariant defined for Weyl semimetals is the Z-type monopole charge \C{} associated with helical surface states~\cite{weng2015weyl,Weyl_Taas,Weyl_experiments,Weyl_exp,fang2016hss,yang2018ideal,chen2016photonic,he2020observation,zhang2018double,zhang2020twofold}. 
Since a Dirac point can be treated as a composition of a pair of Weyl points carrying opposite signs of \C{}, both the monopole charge \C{} and the BSC are expected to vanish in Dirac semimetals. 
However, a recent study gives a strict proof that a \ZZ-type monopole charge \Q{} can be defined for Dirac points when time-reversal (\T{})-glide (\G) symmetry (\Th=${TG}$) is present~\cite{zhang2022local}, and they are associated with anti-parallel HSSs like double/quad-helical surface states (DHSSs/QHSSs)~\cite{yang2014classification,morimoto2014weyl,shiozaki2015z,fang2016hss,gorbar2015surface,cai2020symmetry,cheng2020discovering}. 
Among diverse topological semimetals, Weyl and Dirac semimetals draw the most attention, not only due to their exotic transport properties from the bulk bands, but also due to their topological surface states with fascinating helical shapes.

\begin{figure*}
\centering
\includegraphics[scale=1.0]{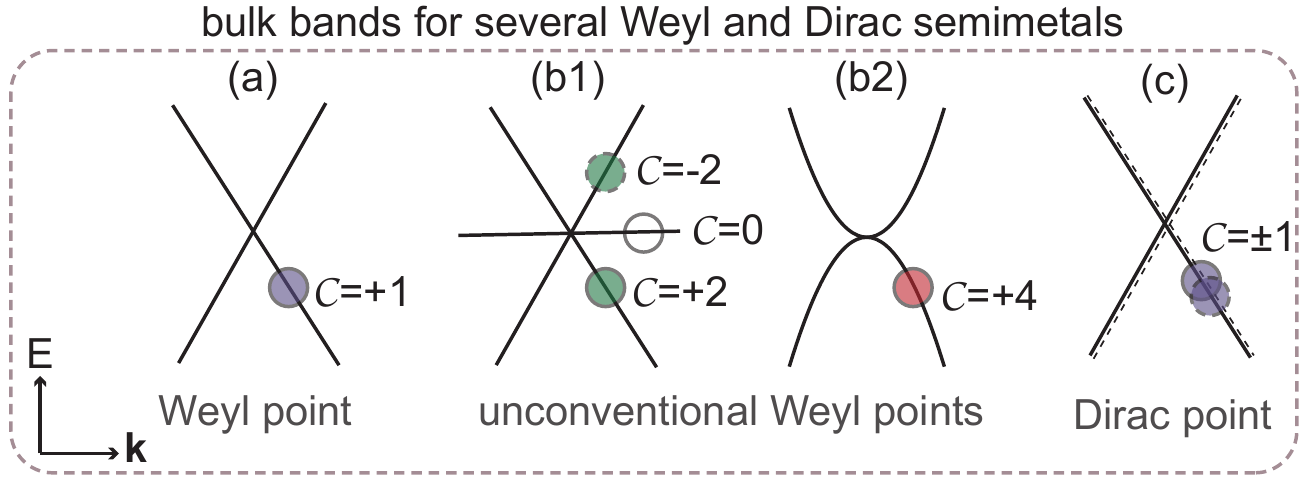}
\caption{Bulk bands for various Weyl points and Dirac points. (a) Linearly dispersive bands for a Weyl point carrying monopole charge \C{} = +1. (b1-b2) Two different kinds of unconventional Weyl points carrying monopole charge of \C{} = +2 and +4, respectively. The threefold degenerate Weyl point in (b1) is also known as spin-1 Weyl point or double Weyl point, while the twofold one in (b2) is also known as twofold quadruple Weyl point. 
(c) A Dirac point carrying a total monopole charge of \C{} = 0, and it is composed of two Weyl points with \C{} = $\pm 1$.}
\label{fig:f1}
\end{figure*}

The simplest case is a Weyl point formed by two linear bands, carrying a unit monopole charge, as the one with \C{} = +1 shown in Fig.~\ref{fig:f1} (a). The low-energy effective Hamiltonian can be written as $H_1(\mathbf{k})=\mathbf{k}\cdot\mathbf{\sigma}$, where $\mathbf{\sigma}$ represent three Pauli matrices.   
Furthermore, unconventional Weyl points carrying monopole charges of $|C|>1$ can be obtained if additional crystalline symmetries are preserved, 
such as the threefold degenerate Weyl point protected by the chiral cubic symmetry shown in Fig.~\ref{fig:f1} (b1). It carries monopole charges of \C{} = +2, 0, $-2$ for each band and is also known as spin-1 Weyl point or double Weyl point~\cite{bradlyn2016beyond,liang2016semimetal,zhu2017emergent,zhang2018double,miao2018observation,tang2017multiple,chang2017unconventional,mai2018topological}. 
The low-energy effective Hamiltonian can be written as $H_2(\mathbf{k})=\mathbf{k}\cdot\mathbf{L}$, where $\mathbf{L}$ represent three spin-1 matrix representations of the rotation generators. 
In addition to the spin-1 Weyl point, chiral cubic symmetry can also give rise to a twofold Weyl point carrying monopole charge of \C{} = +4, $-4$ for each band, as shown in Fig.~\ref{fig:f1} (b2), which is also called twofold quadruple Weyl point~\cite{zhang2020twofold,li2020observation}. 
The low-energy effective Hamiltonian can be written as $H_4(\mathbf{k})=k_xk_yk_z\cdot\sigma_{z}+(k_x^2 +\omega k_y^2 +\omega^2 k_z^2)\cdot\sigma_x$ with $\omega=e^{-\frac{2\pi i}{3}}$.  
Nevertheless, when a system preserves inversion symmetry, Weyl points do not appear while Dirac points composed of a pair of Weyl points with \C{} = +1 and $-1$ appear, as shown in Fig.~\ref{fig:f1} (c). 
The low-energy effective Hamiltonian for Dirac points can be written as 
$H_D(\mathbf{k})=
\big(\begin{smallmatrix}
\mathbf{k}\cdot\mathbf{\sigma} & 0\\
0 & -\mathbf{k}\cdot\mathbf{\sigma}
\end{smallmatrix}\big)$. 

\begin{figure*}
\centering
\includegraphics[scale=0.8]{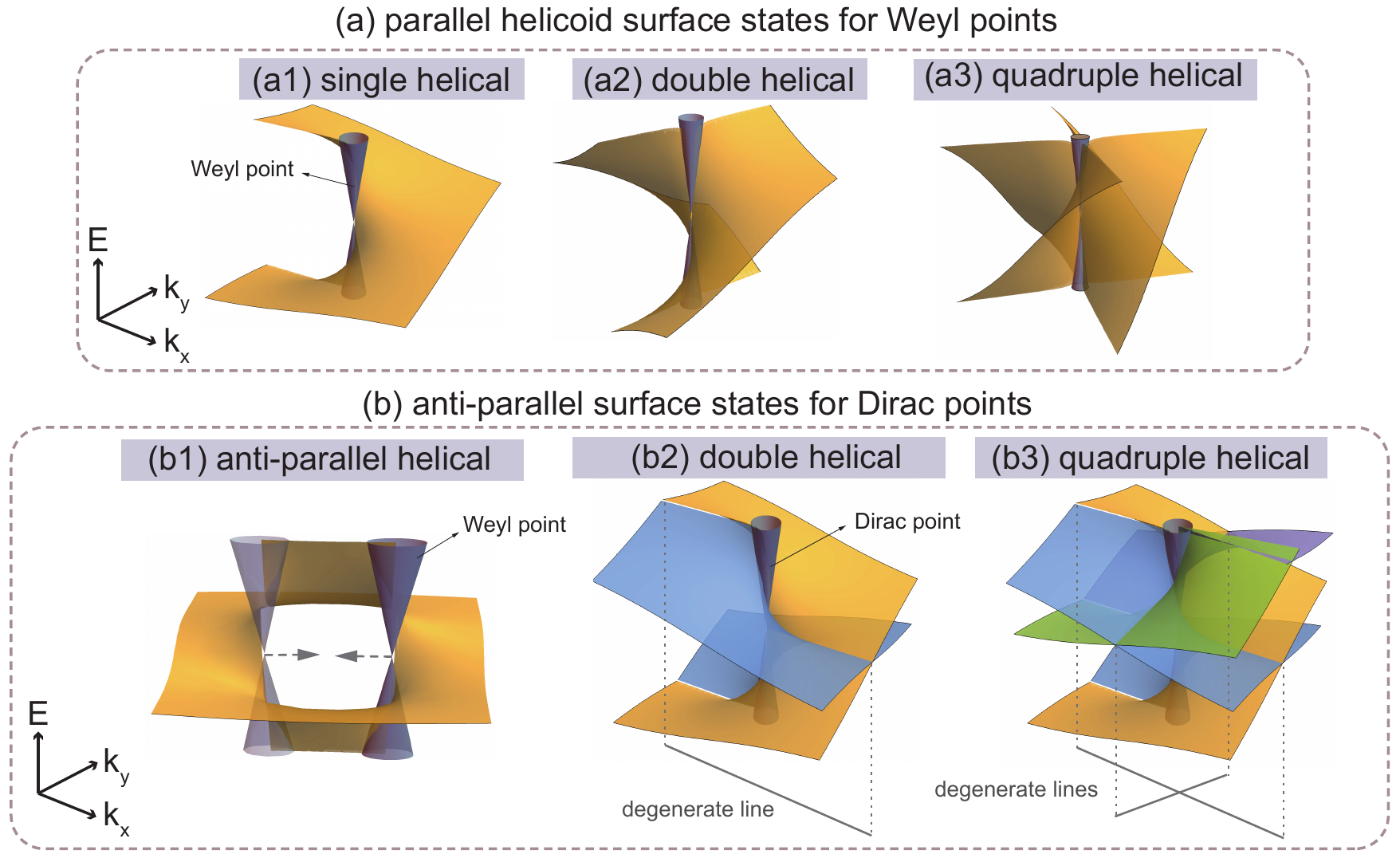}
\caption{(a) Parallel and (b) anti-parallel multi-helical surface states. 
(a1) Single helical surface states for a Weyl point carrying unit monopole charge of \C{} = +1. 
(a2) Parallel double helical surface states for Weyl points carrying monopole charge of \C{} = +2. 
(a3) Parallel quad-helical surface states for Weyl points carrying monopole charge \C{} = +4. 
(b1) Surface states for a pair of Weyl points carrying monopole charge of \C{} = $\pm1$, and it is formed by two helical surface states with different chirality merged together. 
(b2) Anti-parallel double helical surface states for a Dirac point carrying a nonzero \ZZ-type monopole charge \Q{}, where the intersection of the surface states forms a \Th-invariant degenerate line along surface BZ boundary. 
(b3) Anti-parallel quad-helical surface states for Dirac points carrying nonzero \ZZ-type monopole charge \Q{}, where the intersections of the surface states are on the \Th{}-invariant lines protected by two \Th{} symmetries formed by two glide symmetries.  
\label{fig:f2}}
\end{figure*}

Analogous to topological (crystalline) insulators, in which all the topological surface states are composed of surface Dirac cones, the basic unit of the surface states for Weyl and Dirac semimetals can be regarded as the helical surface state~(HSS), as shown in Fig.~\ref{fig:f2} (a1). 
Due to the integer nature of the monopole charge \C{}, HSSs for Weyl points carrying a different monopole charge \C{} will have different multiplicities, 
resulting in parallel surface states with a multi-helical configuration~\cite{bradlyn2016beyond,fang2016hss,chen2016photonic,yang2018ideal,zhang2018double,he2020observation,zhang2020twofold}. 
Figures~\ref{fig:f2} (a2-a3) show the parallel DHSSs and parallel QHSSs for Weyl points carrying monopole charge of \C{} = +2, +4, respectively. 
A pair of isolated Weyl points carrying opposite monopole charge of \C{}$=\pm1$ will preserve a single surface state connecting those two Weyl points, merging into two anti-parallel HSSs with different chirality at the two Weyl points, as shown in Fig.~\ref{fig:f2} (b1). 

However, when those two Weyl points carrying \C{} = $\pm1$ approach each other, a Dirac point will be obtained if there are additional crystalline symmetries (such as inversion and rotation symmetries), and the BSC will no longer necessarily exist due to \C{} = 0. 
Nevertheless, a recent study shows that BSC associated with anti-parallel multi-HSSs can be obtained for Dirac points with a new \ZZ-type monopole charge \Q{}, which is defined in terms of the time-reversal-glide symmetry \Th=\TG{}~\cite{zhang2022local}. 
The study also gives a strict proof that BSC associated with anti-parallel double HSSs can be obtained for \Q{} in both spinless and spinful systems with one \Th{}, while anti-parallel quad-HSSs can be obtained only in spinless systems with two \Th{} symmetries composed of two glide symmetries, as shown in Fig.~\ref{fig:f2} (b2-b3). 

We note that parallel and anti-parallel quad-HSSs are the maximum number of helical surface states that a Weyl and Dirac semimetal can have in crystals. 
Though different studies on parallel and anti-parallel HSSs have been implemented separately in the past, there is no systematically theoretical and numerical research on them, in particular for the multi-HSSs ones. 
In this paper, we will review and discuss the parallel and anti-parallel HSSs associated with Weyl and Dirac points in phonon bands with two new material examples, respectively, and show their BSC associated with different monopole charges. 
One is boron with a chiral cubic crystal structure, in which three kinds of Weyl phonons carrying monopole charge of \C{} = $\pm1$, $\pm2$ and $\pm4$ can be obtained, associated with single, parallel double and parallel quad-HSSs, respectively. 
The other one is silicon with centrosymmetric cubic crystal structure, in which two kinds of Dirac phonons carrying nonzero \ZZ{} monopole charge \Q{} can be obtained, associated with anti-parallel quad-HSSs.

\begin{figure*}
\centering
\includegraphics[scale=1.0]{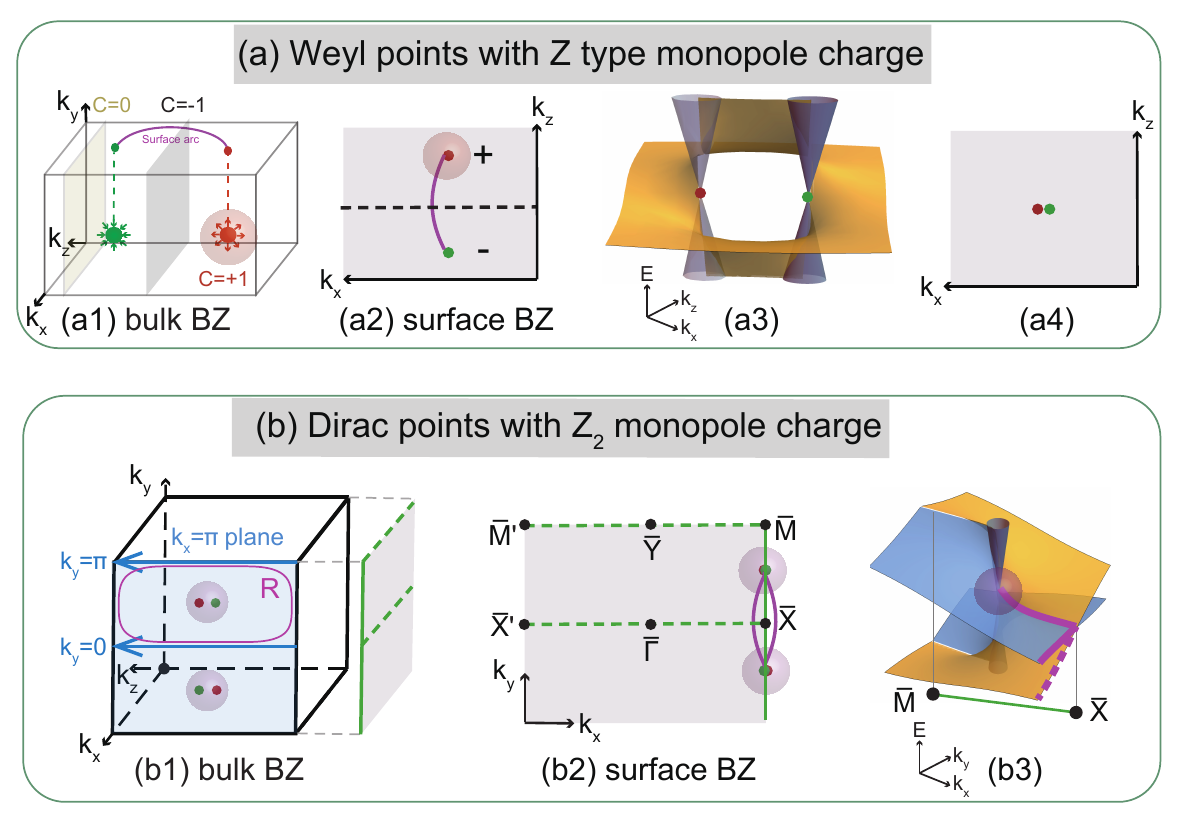}
\caption{Z-type and \ZZ-type monopole charge and their bulk-surface correspondence associated with Weyl and Dirac points. (a1) shows a pair of Weyl points (red and green dots) carrying \C{}= $\pm1$ in the bulk BZ. Since the Z-type monopole charge \C{} can be defined by the Berry curvature integrated either on the sphere enclosing the Weyl point or on the 2D plane away from the Weyl points, showing both the local and global topology of \C{}. 
(a2) Fermi arc connecting two Weyl points carrying  \C{}= $\pm1$ in the surface BZ. 
(a3) Overall configuration for the helical surface states connecting two Weyl points. 
(a4) Fermi arc does not necessarily exist when two Weyl points are projected onto the same point in the surface BZ. 
(b1) Bulk BZ for a system preserving \G$_y$. Two Dirac points marked by green and red dots are located on the $k_x=\pi$ plane (blue plane) satisfying \Th$_y^2=-1$. (b2) Fermi arcs (purple curves) for Dirac points carrying nontrivial \ZZ{} monopole charge \Q{} on the \G$_y$-preserved surface BZ. 
(b3) Double-HSSs contributed by Dirac points with nontrivial \Q{}, which is projected along $\bar{M}$-$\bar{X}$ (satisfying \Th$_y^2=-1$) on the surface BZ. The gray cone is the bulk Dirac bands. Blue and yellow sheets are the anti-parallel HSSs. }
\label{fig:mc}
\end{figure*}

{\section*{Results}}

\subsection*{Z-type monopole charge \C{} associated with Weyl points}
As the topological invariant for Weyl points, the Z-type monopole charge \C{} can be defined by an integral of the Berry curvature in $k$-space, either on a sphere enclosing the Weyl points or on a 2D plane which is away from the Weyl points. 
Thus, $C_n = \frac{1}{2\pi}\int_S\Omega_{n}(\textbf{k})\cdot {\rm d}S$, where Berry curvature for $n^{th}$ band is defined as $\Omega_n(\textbf{k}) = i\langle\nabla_\textbf{k}u_{n\textbf{k}}|\times|\nabla_\textbf{k}u_{n\textbf{k}}\rangle$ in terms of the wavefunction $|u_{n\textbf{k}}\rangle$ and $S$ is the sphere or the 2D plane marked in Figs.~\ref{fig:mc} (a1)-(a2). 
The former definition of \C{} shows the local property of Weyl points, i.e., the monopole charge of \C{}, while the latter definition tells the influence of \C{} to the rest part of the BZ, giving rise to HSSs connecting two Weyl points with opposite monopole charges, as shown in Figs.~\ref{fig:mc} (a2)-(a3). 
An isoenergetic contour of the surface states is the Fermi arc connecting these two Weyl points, as marked by the purple solid line in Figs.~\ref{fig:mc} (a1)-(a2). 
If the Chern number \C{} defined on the 2D plane is nonzero (such as the gray 2D plane in Fig.~\ref{fig:mc} (a1)), there will be the number of |\C{}| Fermi arcs crossing the line, which is a projection of the 2D plane onto the surface BZ (such as the dashed line in Fig.~\ref{fig:mc} (a2)). 
However, the Fermi arc will disappear when those two Weyl points are projected onto the same momentum on the surface BZ or when they meet and become a Dirac point, showing the \C{} = 0 nature of the Dirac points, as shown Fig.~\ref{fig:mc} (a4).




\subsection*{\ZZ-type monopole charge \Q{} associated with Dirac points}

As we mentioned above, the \ZZ-type monopole charge \Q{} can be defined in terms of \Th=\TG{} symmetry for Dirac points. 
Unlike the Z-type monopole charge \C{} characterizing Weyl points, this new monopole charge \Q{} \textit{cannot} be defined as a local quantity and it is associated with anti-parallel HSSs. 

Figure~\ref{fig:mc} (b1) shows the bulk BZ for a system with time-reversal symmetry \T{} and single glide symmetry ${G}_{y}=\{M_y|\frac{1}{2}00\}$, which leads to $\tilde{\Theta}_{y}^2$ = (\T\G$_{y})^2$ = $e^{-ik_x}$. 
Thus, $\tilde{\Theta}_{y}^2=-1$ for the $k_x=\pi$ plane marked by the blue rectangle. 
We note that two \T-related Dirac points carrying nontrivial \Q{} are usually located on high-symmetry lines for systems only has one \G, and at non-TRIM high-symmetry points for systems preserving two \G{} with mutually perpendicular mirror planes. 
In the single glide symmetry case, two Dirac points located on the blue plane shown in Fig.~\ref{fig:mc} (b1) will be projected onto $\bar{M}-\bar{X}$ on the glide-preserving surface BZ shown in Fig.~\ref{fig:mc} (b2). 

At first, the \ZZ-type monopole charge is defined in terms of wavefunctions~(WFs) on a sphere enclosing the Dirac point, which is analogous to the local definition of \C{}. 
Nonetheless, Ref.~\cite{zhang2022local} points out that this definition of \Q{} is not gauge invariant, because it is in fact only defined in terms of the WFs on the circle lying on the $k_x=\pi$ plane enclosing the Dirac point. 
Namely, information on the WFs on this circle is not sufficient to uniquely determine the value of the \ZZ{} invariant \Q{}, and these still exists a gauge transformation which changes \Q{} by unity. 
To make \Q{} well-defined, a rectangle R (marked by purple lines in Fig.~\ref{fig:mc} (b1)) that traverses the half of the $k_x=\pi$ plane should be used, and the Dirac point should be the only gapless point inside of the rectangle.  
By this new definition, \Q{} is equal to the difference of \Th-polarization $P_{\tilde{\Theta}}$ along two blue lines shown in Fig.~\ref{fig:mc} (b1): 

\begin{align}
{Q}&=P_{\tilde{\Theta}}(k_y=\pi)-P_{\tilde{\Theta}}(k_y=0)\pmod{2},
\label{eq:nugamma}
\end{align}
where  $P_{\tilde{\Theta}}(k_y)=\frac{1}{2\pi}(\gamma^+_{\mathrm{L}}-\gamma^-_{\mathrm{L}})$, $\gamma_{\mathrm{L}}$ is the Berry phase along one of the blue lines, and ``$\pm$'' are the positive/negative glide sectors for the Bloch WFs. 
This quantity \Q{} is well-defined only when the system is gapped along the blue lines. 
Thus, although \Q{} is a monopole charge for the Dirac point, we cannot define it on a sphere enclosing the Dirac point, because \Th$_y^2=-1$ holds only on the $k_x=\pi$ plane due to the definition in Eq.~(\ref{eq:nugamma}). 
Instead, \Q{} can be only defined by a global way along the blue lines, and it can be nontrivial when the system is \TG-invariant. 
Furthermore, when we slightly break the time-reversal symmetry, this \ZZ{} invariant \Q{} reduce to the \ZZ{} invariant characterizing the glide-symmetric topological crystalline insulators studied in Refs.~\cite{shiozaki2015z,fang2015new,Kim-nonprimitive,Kim-PC}. 
By using this fact, we can show that a nontrivial value of \Q{} is associated with anti-parallel DHSSs on the ${G}_y$-preserving surface, in both spinless and spinful systems, as shown in Figs.~\ref{fig:mc} (b2)-(b3). 

Now, let us consider a spinless system with \T{} and two \G{} with mutually perpendicular mirror planes, such as ${G}_x=\{M_x|0\frac{1}{2}0\}$ and ${G}_y=\{M_y|\frac{1}{2}00\}$ for space group \#110. \#110 is a body-centered lattice, it will have a similar bulk BZ as the ones shown in Fig.~\ref{fig:f3} (b) and Fig.~\ref{fig:f4} (b). 
Since two glide symmetries will give rise to \Th$_x^2=-1$ at $k_y=\pi$ plane and \Th$_y^2=-1$ at $k_x=\pi$ plane,  \T-related Dirac points carrying nontrivial \Q{} will appear at non-TRIM high-symmetry points, e.g. $P$ and $P^{\prime}$ in the bulk BZ.  
Furthermore, due to the additional $C_{2z}$ symmetry resulting from two \G{} with mutually perpendicular mirror planes in spinless systems, \Q{} can be further simplified in terms of $C_{2z}$ eigenvalues~\cite{Kim-nonprimitive,Kim-PC}:
\begin{align}
(-1)^{Q}=\prod_i \frac{\zeta_i(\Gamma)}{\zeta_i(X)},
\label{eq:Qnew}
\end{align}
where $\zeta_i$ is the eigenvalue of $C_{2z}$ for the $i^{th}$ occupied band. $\Gamma$ and $X$ are two $C_{2z}$-invariant high-symmetry momenta on the $k_z=0$ plane in the space group \#110, which can be replaced by other high-symmetry momenta listed in Tab.~\ref{tab:1} for other space groups preserving two glide symmetries. 
This formula (Eq.~(\ref{eq:Qnew})) clearly shows the global nature of the \ZZ-type monopole charge \Q{}. The BSC between the monopole charge and the DHSS/QHSS is thus established. 

\begin{table}[] 
\begin{tabular}{lccc}
\hline
{\begin{tabular}[c]{@{}l@{}}Space\\ group\end{tabular}} & {\begin{tabular}[c]{@{}c@{}}Two \\ glide mirrors\end{tabular}}         & {Location} & {\begin{tabular}[c]{@{}c@{}}Momenta used\\ for Eq.~(\ref{eq:Qnew}) \end{tabular}} \\ \hline \hline
\#73                               & \{$M_x|\frac{1}{2},\frac{1}{2},0$\}; \{$M_y|\frac{1}{2},0,0$\}                & W    & $\Gamma$, T                                                                       \\
\#110                             & \{$M_x|\frac{1}{2},\frac{1}{2},0$\}; \{$M_y|\frac{1}{2},\frac{1}{2},0$\}  & P     & $\Gamma$, X                                                               \\
\#142                             & \{$M_x|\frac{1}{2},\frac{1}{2},0$\}; \{$M_y|\frac{1}{2},0,0$\}               & P  & $\Gamma$, X                                                                              \\
\#206                             & \{$M_x|\frac{1}{2},\frac{1}{2},0$\}; \{$M_y|\frac{1}{2},0,0$\}               & P    & $\Gamma$, N                                                                       \\
\#228                             & \{$M_x|\frac{1}{2},\frac{3}{4},0$\}; \{$M_y|\frac{3}{4},\frac{1}{2},0$\}   & W    & $\Gamma$, X                                                           \\    
\#230                             &  \{$M_x|\frac{1}{2},\frac{1}{2},0$\}; \{$M_y|\frac{1}{2},0,0$\}              & P  & $\Gamma$, N                                              \\                    
\hline  
\end{tabular} 
\caption{Spinless systems with two \G{} with mutually perpendicular mirror planes, where Dirac points associated with QHSSs can be obtained. Four columns represent the space group number, two glide symmetries which define the \ZZ{} monopole charge \Q{} together with \T{}, the momentum where Dirac points are located and the momenta used for Eq.~(\ref{eq:Qnew}). }
\label{tab:1}    
\end{table}
%



\subsection*{Parallel multi-HSSs for Weyl phonons in boron}
Next we give some material examples where phonon spectra show the multi-HSSs. We begin with the parallel multi-HSSs. 
To obtain the unconventional Weyl points carrying monopole charges of \C{} = $\pm 2$ and $\pm 4$ with $H_2(\mathbf{k})=\mathbf{k}\cdot\mathbf{L}$ and $H_4(\mathbf{k})=k_xk_yk_z\cdot\sigma_{z}+(k_x^2 +\omega k_y^2 +\omega^2 k_z^2)\cdot\sigma_x$, as shown in Figs.~\ref{fig:f1} (b1)-(b2), chiral cubic symmetry, time-reversal symmetry and SU(2) symmetry are necessary. 
From these conditions, we find that boron with space group \#214 is a good candidate for these unconventional Weyl points in phonon band structure, and associated parallel multi-HSSs.

\begin{center}
\begin{figure*}
\includegraphics[scale=0.6]{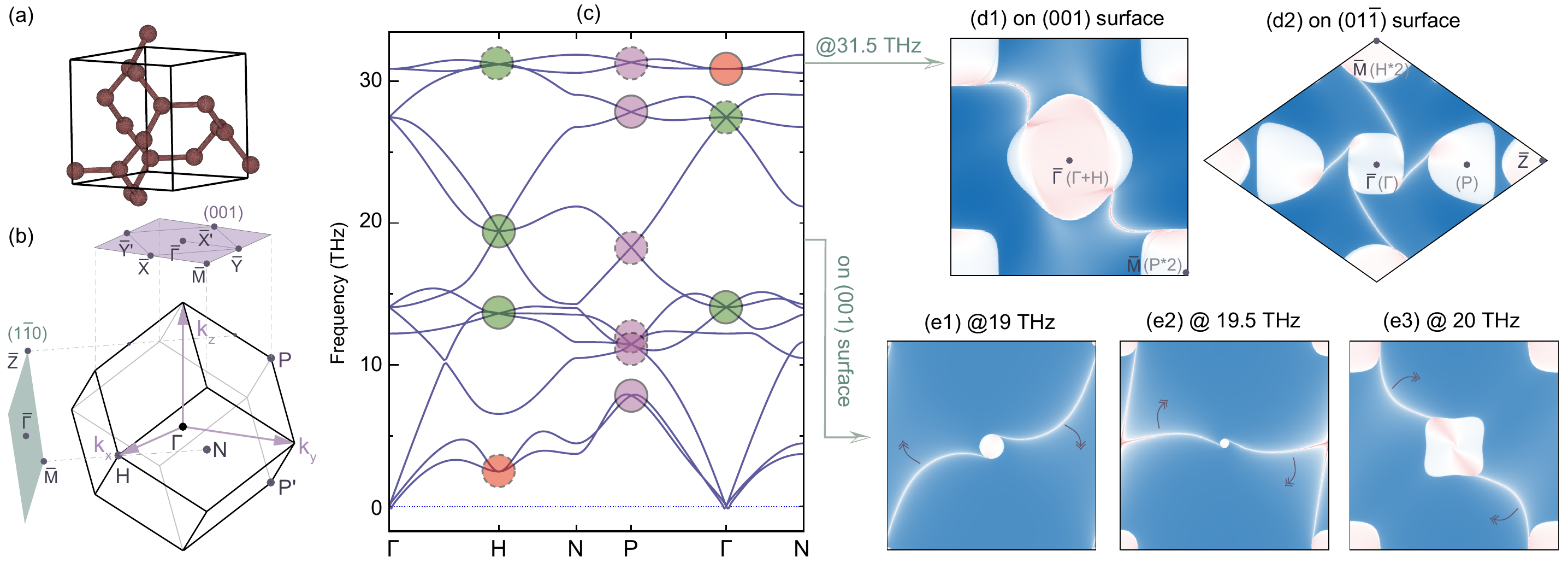}\caption{Weyl phonons in boron with $I4_132$ crystal structure. 
(a) Crystal structure for boron with space group $I4_132$. (b) Bulk, (001) and (1$\bar{1}$0) surface Brillouin zone for boron. 
(c) Phonon spectra of boron, where Weyl points with monopole charge \C{}= +1, +2 and +4 ($-1$, $-2$ and $-4$) are marked by the purple, green and red dots with solid (dashed) circle outside. 
(d1)-(d2) Surface arcs at 31.5 THz on the (001) and (1$\bar{1}$0) surface, respectively, and three Weyl points at $\Gamma$, $H$ and $P$ carrying different monopole charge \C{} will projected onto different momenta in the surface BZ.  
(e1)-(e3) Surface arcs with three different energies around 19.5 THz on the (001) surface, and they rotate around $\bar{\Gamma}$ and $\bar{M}$ as the energy increases. }
\label{fig:f3}
\end{figure*}
\end{center}

Boron with space group \#214 ($I4_132$) belongs to the chiral cubic system, as shown in Fig.~\ref{fig:f3} (a), where three kinds of Weyl points can be obtained in its phonon spectra. 
Due to the body-centered lattice nature of boron, both high-symmetry momenta $H$ and $\Gamma$ will have two kinds of unconventional Weyl points protected by the chiral cubic symmetry, i.e., threefold double Weyl points (\C{}$ = \pm2$) marked by the green dots and twofold quadruple Weyl points (\C{}$ = \pm4$) marked by the red dot, as shown in Figs.~\ref{fig:f3} (b-c). 
The solid and dashed circles outside of the dot represent the positive and negative sign of the monopole charge, respectively. 
Degeneracies at the high-symmetry momentum $P$ are all twofold ones, as marked by the purple dots, and they all correspond to single Weyl points carrying monopole charges of \C{}$ = \pm1$. 
Since the number of parallel HSSs corresponds to the value of the monopole charge $|$\C{}$|$, single, double and quad-HSSs can be obtained in the phonon spectra of boron, as shown in Figs.~\ref{fig:f2} (a1-a3), which can be further demonstrated by the calculation of surface modes.

Figure~\ref{fig:f3} (d1) shows surface mode calculation at 31.5 THz along the (001) surface. Along the calculation of the (001) surface, both the threefold double Weyl phonon at $H$ and the twofold quadruple Weyl phonon at $\Gamma$ will be projected onto $\bar{\Gamma}$ and the conventional Weyl points at $P$ and $P^{\prime}$ will be projected onto $\bar{M}$, as shown in Fig.~\ref{fig:f3} (b). 
Thus, the total monopole charge at $\bar{\Gamma}$ and $\bar{M}$ will be \C{} = $+2$ and $-2$ on the (001) surface, respectively, resulting in two non-contractible surface arcs coming out from the surface BZ center $\bar{\Gamma}$ and connecting to the surface BZ corner $\bar{M}$. Since $C_{2}$ and \T{} symmetries are the only two symmetries left on the (001) surface, two surface arcs appear in a $C_{2}$- and \T-related way. 

Since the threefold double Weyl phonon and twofold quadruple Weyl phonon overlap along $[001]$ direction, the parallel quad-HSS nature of the twofold quadruple Weyl phonon with four surface arcs will also be covered up. 
In order to capture the whole helical configuration of the twofold quadruple Weyl point with \C{} = +4, we also calculate the surface arcs with the same frequency but along (1$\bar{1}$0) surface, as shown in Fig.~\ref{fig:f3} (d2). 
On the ($1\bar{1}0$) surface, the twofold quadruple Weyl point at $\Gamma$ will be projected onto $\bar{\Gamma}$, while the threefold double Weyl point at $H$ will be projected onto a different momentum $\bar{M}$, and the conventional Weyl points at $P$ and $P^{\prime}$ will be projected along $\bar{\Gamma}-\bar{Z}$ directions. 
Such non-overlapping projections along $[1\bar{1}0]$ direction will not only show the parallel quad-HSSs of the twofold quadruple Weyl node at $\bar{\Gamma}$, but also show the parallel double HSSs of the threefold double Weyl node at $\bar{M}$ and the single HSSs from $P$ and $P^{\prime}$. 
Four surface arcs will also display in a $C_{2}$- and \T-related way due to these symmetries on the ($1\bar{1}0$) surface. 

To have an intuitive picture of the parallel HSSs, we also calculate the surface arcs along the (001) surface, with different energies around 19.5 THz, as shown in Figs.~\ref{fig:f3} (e1)-(e3). Along $[001]$ direction, the threefold double Weyl point around 19.5 THz at $\Gamma$ is projected onto $\bar{\Gamma}$ with \C{} = +2 and two conventional Weyl points at $P$ and $P^{\prime}$ will be projected onto $\bar{M}$ with \C{} = $-2$. Thus, double HSSs with two surface arcs connect from $\bar{\Gamma}$ to $\bar{M}$, forming a non-contractible surface arc crossing the whole surface BZ. 
As the energy increases, the double HSSs rotate in the same sense, and this sense of rotation is opposite between $\bar{\Gamma}$ and $\bar{M}$, showing the parallel helical nature of the surface states and different monopole charges of the Weyl points located at $\bar{\Gamma}$ and $\bar{M}$. 
\subsection*{Anti-parallel double HSSs for Dirac points}
Next, we consider a material example for the anti-parallel multi-HSSs. 
In the Dirac semimetal case, where the fourfold degenerate bulk bands are composed of two Weyl points carrying opposite monopole charges of \C{}$=\pm1$, a pair of anti-parallel HSSs are expected to appear. 
However, HSSs expected from the BSC for the monopole charge \C{} will vanish due to the total monopole charge \C{}$=0$ for Dirac points. 
Due to the lack of topological protection, anti-parallel HSSs for Dirac points will hybridize with each other at the intersections  under perturbations, resulting in a surface Dirac cone~\cite{Kargarian8648,le2018dirac} without helical nature.  
Nevertheless, a new BSC associated with \ZZ{} monopole charge \Q{} can be defined for Dirac points when the system preserves time-reversal-glide symmetry \Th{}=\TG{}. 
In contrast to the Z-type monopole charge C, which can be defined both locally and globally, the \ZZ-type monopole charge \Q{} can be only defined in a global way, i.e., it only has a global nature of topology~\cite{fang2016hss,zhang2022local}. 
With the protection of \Q{}, gap opening for the anti-parallel HSSs can be avoided, resulting in anti-parallel double/quad-HSSs, as we discussed in Fig.~\ref{fig:mc}.

Anti-parallel double HSSs consist of two HSSs overlapping on the surface BZ, as shown in Fig.~\ref{fig:f2} (b2), and at their crossing they are expected to become gapped when perturbations are introduced~\cite{kargarian2016surface,kargarian2018deformation,wang2012dirac,liu2014discovery,neupane2014observation,liu2014stable,borisenko2014experimental,guo2017three,le2018dirac,wu2019fragility}. 
However, with the protection of \Th$^2=-1$, a nonzero \ZZ{} monopole charge \Q{} for the Dirac points can be defined~\cite{fang2016hss,zhang2022local}, and a BSC associated with anti-parallel double HSSs can be obtained for both spinless and spinful systems with a single glide mirror symmetry. 
The intersections of the anti-parallel double HSSs are protected by the anti-unitary operator with \Th$^2=-1$, which forms a degenerate line along the BZ boundary, as shown in Fig.~\ref{fig:f2} (b2). 
We note that a Dirac point carrying a nontrivial monopole charge \Q{}  can be transformed into a pair of Weyl points or a nodal ring under any \Th-preserved perturbations. Such a pair of Weyl points, which are known as Weyl dipoles, and the \ZZ{} nodal ring also hold anti-parallel double HSSs, due to the global definition of the monopole charge \Q{}~\cite{zhang2022local}.

\subsection*{Anti-parallel quad-HSSs for Dirac points}

As discussed above, anti-parallel double HSSs appear due to BSC of a nonzero monopole charge \Q{}, but they can be also treated as two HSSs from two Weyl points with opposite chirality. 
If two Dirac points, i.e., two pairs of Weyl points carrying a monopole charge of \C{} = $\pm1$, are projected onto the same momentum in the surface BZ, there will be two pairs of anti-parallel HSSs crossing each other at least four times in the band gap around the projected Dirac points, as shown in Fig.~\ref{fig:f2} (b3). 
Thus, if there are two \Th{} symmetries in a system formed by two \G{} with mutually perpendicular mirror planes, anti-parallel quad-HSSs can be obtained. 
In the anti-parallel quad-HSSs, the two \Th{} symmetries protect the intersections of those two pairs of anti-crossing HSSs from opening a gap along the \Th-invariant lines in the surface BZ.

However, in \textit{spinful} systems with two glide symmetries, due to the presence of other additional bulk degeneracies at time-reversal-invariant momenta~(TRIM) located on the blue lines in Fig.~\ref{fig:mc} (b1), the definition (\ref{eq:nugamma}) of \Q{} for the Dirac point is ill-defined. Therefore, the BSC for \Q{} with anti-parallel quad-HSSs can be obtained only in \textit{spinless} systems with two glide symmetries; in such systems, two \T-related Dirac points with nontrivial \Q{} are at the high-symmetry momenta, which are not TRIM. 
Thus, a face/body-centered lattice with two \G{} with mutually perpendicular mirror planes are the necessary conditions to obtain Dirac points associated with anti-parallel quad-HSSs. In the next session, we will use silicon in a body-centered cubic structure as an example to show the anti-parallel quad-HSSs associated in its phonon spectra.

\subsection*{Anti-parallel multi-HSSs for Dirac phonons in silicon}
\begin{center}
\begin{figure*}
\includegraphics[scale=0.8]{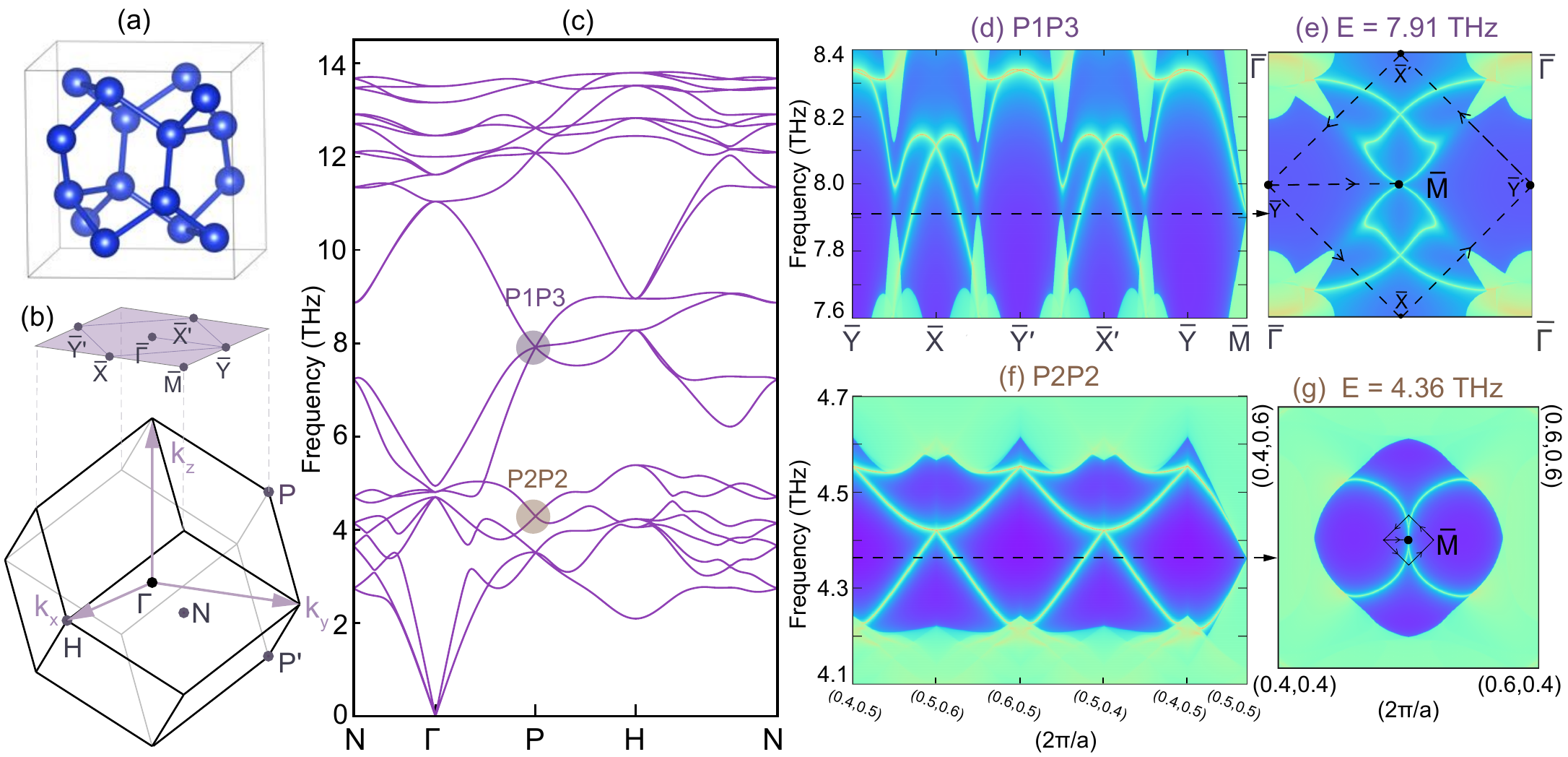}\caption{Dirac phonon in silicon with space group $Ia\bar{3}$.
(a) Crystal structure, (b) bulk BZ and (001) surface BZ and (c) phonon bands for Si with space group $Ia\bar{3}$. $P_{1}P_3$ and $P_2P_2$ marked by colored dots in (c) are both Dirac points but with different irreps. (d)-(e) are the calculated surface states and arcs for two Dirac points located at $P$ and $P^{\prime}$ with irreps of $P_1P_3$. (f)-(g) are the calculated surface states and arcs for two Dirac points located at $P$ and $P^{\prime}$ with irreps of $P_2P_2$.}
\label{fig:f4}
\end{figure*}
\end{center}

In order to have an intuitive comprehension on the anti-parallel quad-HSSs, we propose an experimentally synthesized material candidate, i.e., silicon with \#206~($Ia\bar{3}$) having two kinds of Dirac phonons in its phonon band structure, to show their topological properties. 
Figures~\ref{fig:f4} (a) and (b) are the crystal structure and BZ for Si with \#206~\cite{wosylus2009crystal}, which is an experimentally synthesized phase for silicon, with two glide mirrors ${G}_{x}=\{M_x|0\frac{1}{2}0 \}$ and ${G}_{y}=\{M_y|\frac{1}{2}00\}$ with their mirror planes perpendicular with each other. 
The body-centered structure of Si gives rise to two non-TRIM high-symmetry points $P$ and $P^{\prime}$ related by \T. 
All the phonon bands at $P$ and $P^{\prime}$ are fourfold degenerate, with either of the two kinds of irreducible representations (irreps), i.e., $P_2P_2$ and $P_1P_3$ shown in Fig.~\ref{fig:f4} (c). Both of them are Dirac points carrying a nonzero monopole charge \Q{}. 

To obtain the anti-parallel quad-HSSs for both of the Dirac points with different irreps, surface state calculations with different energy windows are implemented along the (001) surface. 
There are two \Th{} symmetries on the (001) surface BZ with \Th$_x^2$ = (\TG$_x)^2$ = $-1$ and \Th$_y^2$ = (\TG$_y)^2$ = $-1$ along the BZ boundaries $\bar{M}-\bar{X}$ and $\bar{M}-\bar{Y}$, and Dirac points at $P$ and $P^{\prime}$ will be projected onto the corner point $\bar{M}$, as shown in Fig.~\ref{fig:f4} (b). 
Surface state calculations around the Dirac points with irreps $P_1P_3$ (marked in Fig.~\ref{fig:f4} (c)) are shown in Figs.~\ref{fig:f4} (d) and (e), where the former one follows the $k$-path marked by dashed lines in the latter one, and the latter one with the iso-energy counter marked by the black dashed line in the former one. 
Due to \Th$^2=-1$ along the surface BZ boundaries of the (001) surface, surface states will be doubly degenerate, showing Kramers'-like degeneracy feature, as shown in Fig.~\ref{fig:f4} (d), where the surface states are doubly degenerate at $\bar{Y}$, $\bar{X}$, $\bar{X^{\prime}}$, $\bar{Y^{\prime}}$ and along $\bar{Y}-\bar{M}$ . 
Along the iso-energy contour at 7.91 THz shown in Fig.~\ref{fig:f4} (e), four anti-parallel quad-HSSs cross with each other along $\bar{X}-\bar{M}-\bar{X^{\prime}}$, and the intersection is protected by \Th$_x$ from opening a gap. 
Thus, though the surface state calculation along the $k$-path enclosing the projected Dirac points, which crosses the surface BZ boundaries, the global topology of \Q{} is revealed as the anti-parallel quad-HSSs along the $k$-path. 

Calculations on surface states around the Dirac point with irreps $P_2P_2$ (marked in Fig.~\ref{fig:f4} (c)) are shown in Figs.~\ref{fig:f4} (f) and (g). By following a smaller $k$-path enclosing $\bar{M}$ on the (001) surface BZ shown in Fig.~\ref{fig:f4} (g), there are also two groups of anti-HSSs around $\bar{M}$, protected by \Th$_x$ and \Th$_y$. 
Thus, we claim that anti-parallel quad-HSSs can originate from both of the Dirac points with irreps of $P_2P_2$ and $P_1P_3$ in the phonon spectra of Si.

\section*{Discussion}

As two of the most intriguing topological surface states, both parallel and anti-parallel helical surfaces are composed of helical surface states, but they reflect different topologies of the system. 
In this work, we systematically study these two types of (multi-)helical surface states both theoretically and numerically with two material examples. 
The former parallel helical surfaces show both the local and global nature of Z-type monopole charge \C{} associated with Weyl points, while the latter anti-parallel helical surface states show the global nature of \ZZ-type monopole charge \Q{} associated with Dirac points. 
Thus, the number of the parallel helical surfaces is equal to $|$\C{}$|$ in Weyl semimetals, while the anti-parallel ones show the number of the time-reversal-glide symmetries in Dirac semimetals. 
Considering the restriction of both onsite symmetries and crystalline symmetries, quad-helical surface states are the maximal ones that both parallel and anti-parallel surface states can have.

%
%
%
%
%
%
%
%
%
%
%
%
%

\bibliography{reference}
\newpage{}

\end{document}